\documentclass[12pt]{article}
\begin{document}

\baselineskip 0.75cm
\topmargin -0.6in
\oddsidemargin -0.1in

\let\ni=\noindent

\renewcommand{\thefootnote}{\fnsymbol{footnote}}

\newcommand{\CKM}{Cabibbo--Kobayashi--Maskawa }

\pagestyle {plain}

\setcounter{page}{1}

\pagestyle{empty}

~~~

\begin{flushright}
IFT--01/03
\end{flushright}

\vspace{0.3cm}

{\large\centerline{\bf Prediction of LSND effect as 
a "\,$\!$sterile"~perturbation }}

\vspace{0.2cm}

{\large\centerline{\bf of the bimaximal texture for three active 
neutrinos{\footnote {Supported in part by the Polish State Committee 
for Scientific Research (KBN).}}}}

\vspace{0.3cm}

{\centerline {\sc Wojciech Kr\'{o}likowski}}

\vspace{0.3cm}

{\centerline {\it Institute of Theoretical Physics, Warsaw University}}

{\centerline {\it Ho\.{z}a 69,~~PL--00--681 Warszawa, ~Poland}}

\vspace{1.8cm}

{\centerline{\bf Abstract}}

\vspace{0.3cm}

As a contribution to the hypothesis of mixing of three active neutrinos with, 
at least, one sterile neutrino, we report on a simple $4\times 4$ texture 
whose 
$3\times3$ part arises from the popular bimaximal texture for three active 
neutrinos  $\nu_e\,, \,\nu_\mu\,, \,\nu_\tau$, where $c_{12}=1/\sqrt{2} = 
s_{12}$, $ c_{23} = 1/\sqrt{2} = s_{23}$ and $ s_{13} = 0$. Such a $3\times 3
$ bimaximal texture is perturbed through a rotation in the 14 plane, where 
$\nu_4 $ is the extra neutrino mass state induced by the sterile neutrino 
$\nu_s $ which becomes responsible for the LSND effect. Then, with $m^2_1
\simeq m^2_2$ we predict that $\sin^2 2\theta_{\rm atm} = \frac{1}{2}(1+ 
c^2_{14}) \sim 0.95$ and $\sin^2 2\theta_{\rm LSND} = \frac{1}{2}s^4_{14} \sim 
5\times10^{-3}$, and in addition $ \Delta m^2_{\rm atm} = \Delta m^2_{32}$ 
and $ \Delta m^2_{\rm LSND} = |\Delta m^2_{41}|$,  where $c^2_{14} = \sin^2 
2\theta_{\rm sol} \sim 0.9$ and $\Delta m^2_{21} = \Delta m^2_{\rm sol} \sim
10^{-7}\;{\rm eV}^2$ if {\it e.g.} the LOW solar solution is applied. 

\vspace{0.6cm}

\ni PACS numbers: 12.15.Ff , 14.60.Pq , 12.15.Hh .

\vspace{0.8cm}

\ni February 2001

\vfill\eject

~~~
\pagestyle {plain}

\setcounter{page}{1}

\vspace{0.2cm}

The present status of experimental data for atmospheric $\nu_\mu $'s as 
well as solar $ \nu_e $'s favours oscillations between three conventional 
neutrinos $ \nu_e\,, \,\nu_\mu\,, \,\nu_\tau $ only [1]. However, the problem 
of the third neutrino mass-square difference, related to the possible LSND 
effect for accelerator $\nu_\mu $'s, is still actual [2], stimulating a 
further discussion about mixing of these three active neutrinos  with, at 
least, one hypothetical sterile neutrino $ \nu_s $ (although such a sterile 
neutrino is not necessarily able to explain the LSND effect [3]). As a 
contribution to this discussion, we report in this note on a simple 
$4\times 4$ texture for three active and  one sterile neutrinos, 
$ \nu_e\,, \,\nu_\mu\,, \,\nu_\tau $ and $ \nu_s $, whose $3\times 3$ 
part arises from the popular bimaximal texture [4] working {\it grosso modo} 
in a satisfactory way for solar $\nu_e $'s and atmospheric $\nu_\mu $'s if 
the LSND effect is ignored. Such a $3\times 3$  bimaximal texture is 
perturbed [5] by the sterile neutrino $\nu_s$ inducing one extra neutrino 
mass state $\nu_4$ and so, becoming responsible for the possible LSND effect. 
In fact, with the use of our $4\times 4$ texture we predict that 
$\sin^2 2\theta_{\rm LSND} = \frac{1}{2} s^4_{14}$ and 
$ \Delta m^2_{\rm LSND} = |\Delta m^2_{41}|$, while 
$\sin^2 2\theta_{\rm sol} = c^2_{14}$ and 
$\Delta m^2_{\rm sol} = \Delta m^2_{21}$ as well as  
$\sin^2 2\theta_{\rm atm} = \frac{1}{2}(1+ c^2_{14})$ and 
$ \Delta m^2_{\rm atm} = \Delta m^2_{32}$, if $m^2_1 \simeq m^2_2$ 
(and both are different enough from $m^2_3$ and $m^2_4$). Here, 
$c^2_{14} \sim 0.9$ and $\Delta m^2_{21} \sim 10^{-7} \,{\rm eV}^2$ if 
{\it e.g.} the LOW solar solution [1] is accepted; then we predict 
$\sin^2 2 \theta_{\rm atm} \sim 0.95$ and 
$\sin^2 2\theta_{\rm LSND} \sim 5\times10^{-3} $. 
 
In the popular $3\times 3$ bimaximal texture the mixing matrix has the form [4]

%Eq. 1
\begin{equation} 
U^{(3)} =   \left( \begin{array}{ccc} 1/\sqrt{2} & 1/\sqrt{2} & 0 \\
- 1/{2}\, & \;\,1/2 & 1/\sqrt{2} \\ \;\,1/2 & -1/2\, 
& 1/\sqrt{2}\end{array} \right) \;.
\end{equation} 

\ni Such a form corresponds to $ c_{12} = 1/\sqrt{2} = s_{12}$, 
$ c_{23} = 1/\sqrt{2} = s_{23}$ and $ s_{13} = 0$ in the notation 
used for a generic \CKM$\!\!$--type matrix [6] (if the LSND effect 
is ignored, the upper bound $|s_{13}| \stackrel{<}{\sim} 0.1 $ 
follows from the negative result of Chooz reactor experiment [7]). 
Going out from the form (1), we propose in the $4\times 4$ texture 
the following mixing matrix:

%Eq. 2
\begin{equation} 
U\! = \! \left( \!\begin{array}{cccc} \, & \, & \, & 0 \\ \, & 
U^{(3)} & \, & 0 \\  \, & \, & \, &  0  \\ 0  & 0  & 0  & 
1 \end{array}\! \right) \left( \!\begin{array}{cccc} \;\,c_{14}  & 
0  & 0 & s_{14} \\ \;\,0 & 1 & 0 & 0 \\ \;\,0 & 0 & 1 & 0 \\ -s_{14} & 
0 & 0 & c_{14} \end{array} \!\right)\! = \!\left( \!\begin{array}{cccc}
 \;\,c_{14}/\sqrt{2} & \,1/\sqrt{2} & 0 & \;\,s_{14}/\sqrt{2} \\
 -\!c_{14}/2\, & \;\,1/2 & 1/\sqrt{2} & -s_{14}/2\, \\ \;\,c_{14}/2 &
 \!\!-1/2 & 1/\sqrt{2}  & \;\,s_{14}/2 \\ -s_{14} & \;\,0 & 0 & \;\,c_{14}
 \end{array}\! \right)
\end{equation} 

\vspace{-0.2cm}

\ni with $ c_{14} = \cos \theta_{14}$ and $ s_{14} = \sin \theta_{14}$ 
(note that in Eq. (2) only $s_{12}$, $s_{23}$ and $s_{14}$ of all $s_{ij}$ 
with $i,j = 1,2,3,4 \,,\;i<j$ are nonzero).The unitary transformation 
describing the mixing of four neutrinos 
$\nu_\alpha = \nu_e\,, \,\nu_\mu\,, \,\nu_\tau \,,\, \nu_s $ is inverse to 
the form 

\vspace{-0.2cm}

%Eq. 3
\begin{equation}
\nu_\alpha = \sum_i U_{\alpha i} \nu_i \;,
\end{equation}

\ni where $\nu_i = \nu_1\,, \,\nu_2\,, \,\nu_3\,,\, \nu_4 $ denote four 
massive neutrino states carrying the masses 
$m_i = m_1\,, \,m_2\,, \,m_3\,,\, m_4 $. Here, 
$U = \left(U_{\alpha i } \right)\,,\,\alpha = e\,,\,\mu\,,\,\tau\,,\,s$ 
and $i = 1,2,3,4$. Of course, $U^\dagger = U^{-1}$ and also $U^* = U$, so 
that a tiny CP violation is ignored.

In the representation, where the mass matrix of three charged leptons 
$e^- \,,\, \mu^- \,,\, \tau^-  $ is diagonal, the $4 \times 4$ neutrino 
mixing matrix $ U $ is at the same time the diagonalizing matrix for the 
$4 \times 4$ neutrino mass matrix $M = \left(M_{\alpha \beta } \right) $:

\vspace{-0.2cm}

%Eq. 4
\begin{equation} 
U^\dagger M U = {\rm diag}(m_1\,,\,m_2\,,\,m_3\,,\,m_4)\;,
\end{equation} 

\ni where by definition $ m^2_1 \leq m^2_2 \leq m^2_3$ and {\it either} 
$ m^2_4 \leq m^2_1$ {\it or} $ m^2_3 \leq m^2_4$. Then, due to the 
formula $ M_{\alpha \beta} = \sum_i U_{\alpha i} m_i U^*_{\beta i}$ we obtain

\vspace{-0.2cm}

%Eq. 5
\begin{eqnarray} 
M_{e e} & = & \frac{1}{2}\left( c^2_{14} m_1 + s^2_{14} m_4 + m_2\right)  
\;, \nonumber \\
M_{e \mu} & = & - M_{e \tau} = - \frac{1}{2\sqrt{2}} 
\left( c^2_{14} m_1 + s^2_{14} m_4 - m_2 \right)\;, \nonumber \\ 
M_{\mu \mu} & = & M_{\tau \tau} = \frac{1}{2} \left[\frac{1}{2} 
\left( c^2_{14} m_1 + s^2_{14} m_4 + m_2\right) + m_3 \right] = 
M_{e e} + M_{\mu \tau}\;, \nonumber \\ 
M_{\mu \tau} & = & - \frac{1}{2} \left[\frac{1}{2} \left( c^2_{14} m_1 + 
s^2_{14} m_4 + m_2\right) - m_3 \right] \;, \nonumber \\ 
M_{e s} & = & - \frac{1}{\sqrt{2}}\, c_{14}\, s_{14} \left( m_1- m_4 \right) 
\;, \nonumber \\ 
M_{\mu s} & = & - M_{\tau s} = \frac{1}{2}\,c_{14}\, s_{14}
\left(m_1 - m_4 \right) = - \frac{1}{\sqrt{2}} M_{e s}\;, \nonumber \\  
M_{s s} & = & s^2_{14} m_1 + c^2_{14} m_4 \;.
\end{eqnarray}

\vspace{-0.2cm}

\ni Of course, $M^\dagger = M^{-1}$ and also $M^* = M$. From 
Eqs. (5) we find that 

\vspace{-0.2cm}

%Eq. 6
\begin{eqnarray} 
m_{1,4}\;\,{\rm or}\;\,m_{4,1} & = & \frac{M_{e e} - M_{e \mu}\sqrt{2} + 
M_{s s}}{2} \pm \sqrt{ \left( \frac{M_{e e} - M_{e \mu}\sqrt{2} - 
M_{s s}}{2} \right)^{2} + 2 M^2_{e s} }\,, \nonumber \\ 
m_{2} & = & M_{e e} + M_{e \mu}\sqrt{2}\;\;\;\;,\;\;\;\; 
m_3\;\; = \;\;M_{\mu \mu} + M_{ \mu \tau}
\end{eqnarray}

\ni if $m_4 \leq m_1$ or $m_1 \leq m_4$, respectively, and 

\vspace{-0.2cm}

%Eq. 7
\begin{equation} 
(2c_{14}s_{14})^2 = \frac{8M^2_{e s}}{(M_{e e} - M_{e \mu}\sqrt{2} - 
M_{s s})^2 + 8 M^2_{e s}}\; .
\end{equation}

\ni Obviously, $ m_1 + m_2 + m_3 + m_4 = M_{e e} + M_{ \mu \mu} +  
M_{\tau \tau} + M_{s s}$ as $  M_{e e} = M_{\mu \mu} - M_{\mu \tau}$ 
and $ M_{\mu \mu} = M_{\tau \tau}$.

Due to the mixing of four neutrino fields described in Eq. (3), neutrino 
states mix according to the form

%rownanie 8
\begin{equation} 
|\nu_\alpha \rangle = \sum_i U^*_{\alpha i} |\nu_i \rangle\,.
\end{equation}

\ni This implies the following familiar formulae for probabilities of 
neutrino oscillations $ \nu_\alpha \rightarrow \nu_\beta $ on the energy shell:

%Eq. 9
\begin{equation} 
P(\nu_\alpha \rightarrow \nu_\beta) = 
|\langle \nu_\beta| e^{i PL} |\nu_\alpha  
\rangle |^2 = \delta _{\beta \alpha} - 4\sum_{j>i} 
U^*_{\beta j} U_{\beta i} U_{\alpha j} U^*_{\alpha i} \sin^2 x_{ji} \;,
\end{equation}

\ni valid if the quartic product $ U^*_{\beta j} U_{\beta i} 
U_{\alpha j} U^*_{\alpha i} $ is real, what is certainly true 
when a tiny CP violation is ignored (then $U^*_{\alpha i} = U_{\alpha i} $). 
Here,

%Eq. 10
\begin{equation} 
x_{ji} = 1.27 \frac{\Delta m^2_{ji} L}{E} \;\;,\;\;  \Delta m^2_{ji} = 
m^2_j - m^2_i
\end{equation}

\ni with $\Delta m^2_{ji}$, $L$ and $E$ measured in eV$^2$, km and GeV, 
respectively ($L$ and $E$ denote the experimental baseline and neutrino 
energy, while $ p_i = \sqrt{E^2 - m_i^2} \simeq E -  m^2_i/2E $ are 
eigenstates of the neutrino momentum $P$).

With the use of oscillation formulae (9), the proposal (2) for 
the $ 4\times 4$ neutrino mixing matrix leads to the probabilities

%Eq. 11
\begin{eqnarray}
P(\nu_e \rightarrow \nu_e) & \simeq & 1 -  c^2_{14} \sin^2 x_{21} -  
\left(1+ c^2_{14}\right) s^2_{14} \sin^2 x _{41} \;, \nonumber \\
P( \nu_\mu \rightarrow \nu_\mu) & = & P( \nu_\tau \rightarrow 
\nu_\tau)  \simeq 1 - \frac{1}{4} c^2_{14} \sin^2 x _{21} - 
\frac{1}{2}(1 + c^2_{14}) \left( \sin^2 x _{32} +\frac{1}{2} s_{14}^2 
\sin^2 x _{41}\right)  \nonumber \\ & & - \frac{1}{2} s_{14}^2 
\sin^2 x _{43} \;, \nonumber \\ 
P( \nu_\mu \rightarrow \nu_e) & = & P( \nu_\tau \rightarrow \nu_e) 
\simeq \frac{1}{2} \left( c^2_{14} \sin^2 x_{21} + s^4_{14}  
\sin^2 x _{41}\right) \;, \nonumber \\
P( \nu_\mu \rightarrow \nu_\tau) & \simeq & - 
\frac{1}{4} c^2_{14} \sin^2 x_{21} + \frac{1}{2} (1+ c^2_{14}) 
\left( \sin^2 x _{32} - \frac{1}{2} s_{14}^2 \sin^2 
x _{41} \right) \nonumber \\ & &  + \frac{1}{2} s_{14}^2 \sin^2 x _{43} 
\end{eqnarray}

\ni in the approximation, where $ m^2_1 \simeq m^2_2 $ (and both are 
different enough from $ m^2_3 $ and $m^2_4 $). The probabilities 
involving the sterile neutrino $\nu_s $ read:

%Eq. 12
\begin{eqnarray}
P(\nu_\mu \rightarrow \nu_s) & = & P(\nu_\tau \rightarrow \nu_s) = 
(c_{14} s_{14})^2 \sin^2 x _{41} \;, \nonumber \\
P(\nu_e \rightarrow \nu_s) & = & 2 (c_{14} s_{14})^2 
\sin^2x _{41} \;, \nonumber \\ 
P(\nu_s \rightarrow \nu_s) & = & 1 - (2 c_{14} s_{14})^2 
\sin^2 \!x _{41}\;.
\end{eqnarray}

If $\Delta m^2_{21} \ll |\Delta m^2_{41}|$ ({\it i.e.}, 
$x_{21} \ll |x_{41}|$) and

%Eq. 13
\begin{equation} 
\Delta m^2_{21} = \Delta m^2_{\rm sol} \sim 10^{-7} \;{\rm eV}^2  \;,
\end{equation} 

\ni then, under the conditions of solar experiments the first Eq. (11) gives 

%Eq. 14
\begin{equation} 
P(\nu_e \rightarrow \nu_e)_{\rm sol} \simeq 1 -  c^2_{14} 
\sin^2 (x _{21})_{\rm sol} - \frac{1}{2} (1+c_{14}^2) s_{14}^2  
\end{equation} 

\ni with the estimate

%Eq. 15
\begin{equation} 
c^2_{14} = \sin^2 2\theta_{\rm sol} \sim 0.9\;\,,\,\;\frac{1}{2} 
(1 + c^2_{14})
s^2_{14} \sim 0.095\,.  
\end{equation} 

\ni In Eqs. (13) and (15) the LOW solar solution [1,8] is used. Note that

%Eq. 16
\begin{equation} 
P(\nu_e \rightarrow \nu_e)_{\rm sol} \simeq 1 - 
P(\nu_e \rightarrow \nu_\mu)_{\rm sol} - 
P(\nu_e \rightarrow \nu_\tau)_{\rm sol} - (c_{14} s_{14})^2
\end{equation} 

\ni with $(c_{14} s_{14})^2 \sim 0.09$.

If $\Delta m^2_{21} \ll \Delta m^2_{32} 
\ll |\Delta m^2_{41}|\,,\, 
|\Delta m^2_{43}|$ ({\it i.e.}, $x_{21} \ll x_{32} 
\ll |x_{41}| \,,\, |x_{43}|$) and 

%Eq. 17
\begin{equation} 
\Delta m^2_{32} = \Delta m^2_{\rm atm} \sim 3 \times10^{-3}\;{\rm eV}^2 \;, 
\end{equation} 

\ni then for atmospheric experiments the second Eq. (11) leads to

%Eq. 18
\begin{equation} 
P(\nu_\mu \rightarrow \nu_\mu)_{\rm atm} \simeq  1 - 
\frac{1}{2}(1 + c^2_{14}) \sin^2 (x _{32})_{\rm atm} - 
\frac{1}{8} (3+c^2_{14}) s_{14}^2
\end{equation} 

\ni with the prediction

%Eq. 19
\begin{equation} 
\sin^2 2\theta_{\rm atm} = \frac{1}{2}(1 + c^2_{14}) \sim 0.95\;\;,\;\, 
\frac{1}{8} (3+c^2_{14}) s_{14}^2 \sim 0.049 
\end{equation} 

\ni following from the value (15) of $c^2_{14}$. Notice that

%Eq. 20
\begin{equation} 
P(\nu_\mu \rightarrow \nu_\mu)_{\rm atm} 
\simeq 1 - P( \nu_\mu \rightarrow \nu_\tau)_{\rm atm} - 
\frac{1}{4}(1 + c_{14}^2) s_{14}^2 
\end{equation} 

\ni with $(1 + c^2_{14}) s_{14}^2/4 \sim 0.048$.

Eventually, if $\Delta m^2_{21} \ll |\Delta m^2_{41}|$ 
({\it i.e.}, $x_{21} \ll  |x_{41}| $) and 

%Eq. 21
\begin{equation} 
|\Delta m^2_{41}| = \Delta m^2_{\rm LSND} \sim 1 \;{\rm eV}^2 \;,
\end{equation} 

\ni then for the LSND accelerator experiment the third Eq. (11) implies

%Eq. 22
\begin{equation} 
P( \nu_\mu \rightarrow \nu_e)_{\rm LSND}  \simeq \frac{1}{2}s^4_{14} 
\sin^2 (x _{41})_{\rm LSND}
\end{equation} 

\ni with the prediction

%Eq. 23
\begin{equation} 
\sin^2 2\theta_{\rm LSND} = \frac{1}{2} s^4_{14} \sim 5\times 10^{-3}
\end{equation} 

\ni inferred from the value (15) of $ c^2_{14}$. Such a prediction for 
$\sin^2 2\theta_{\rm LSND}$ is not inconsistent with the estimate 
$\Delta m^2_{\rm LSND} \sim 1 \;{\rm eV}^2$ [2]. Note that

%Eq. 24
\begin{equation} 
P(\nu_\mu\, \rightarrow\, \nu_e)_{\rm LSND} \simeq \frac{1}{2} 
\left( \frac{s_{14}}{c_{14}} \right)^2  P( \nu_\mu \rightarrow 
\nu_s)_{\rm LSND}
\end{equation} 

\ni with $\frac{1}{2}(s_{14}/c_{14})^2 \sim 0.062$.

Concluding, we can say that Eqs. (14), (18) and (22) are consistent with 
solar, atmo\-spher\-ic and LSND experiments. All three depend on one common 
correlating parameter $c^2_{14}$, implying 
$c^2_{14} = \sin^2 2\theta_{\rm sol} 
\sim 0.9$, $\sin^2 2\theta_{\rm atm} = 
\frac{1}{2}(1 + c^2_{14}) \sim 0.95$ and 
$\sin^2 2\theta_{\rm LSND} = \frac{1}{2}s^4_{14} \sim 5\times 10^{-3}$. 
They depend also on three different mass-square scales 
$\Delta m^2_{21} = \Delta m^2_{\rm sol} \sim 10^{-7} \,{\rm eV}^2$, 
$\Delta m^2_{32} = \Delta m^2_{\rm atm} \sim 3\times 10^{-3} \,{\rm eV}^2 $ 
and $|\Delta m^2_{41}| = \Delta m^2_{\rm LSND} \sim 1 \,{\rm eV}^2$. Here, 
the LOW solar solution [1,8] is accepted. Note that in Eqs. (14) and (18) 
there are constant terms which modify moderately the usual two--flavor 
formulae. Any LSND--type accelerator project, in contrast to the solar and 
atmospheric experiments, investigates a small $\nu_\mu \rightarrow \nu_e$ 
oscillation effect caused possibly by the sterile neutrino $\nu_s$. Thus, 
this effect (if it exists) plays the role of a small "\,$\!$sterile" 
perturbation of the basic bimaximal texture for three active neutrinos 
$ \nu_e\,, \,\nu_\mu\,, \,\nu_\tau $. Of course, if $s_{14}$ were zero, 
the LSND effect would not exist and both solar $\nu_e \rightarrow \nu_e$ 
and atmospheric $\nu_\mu \rightarrow \nu_\mu $ oscillations would be maximal. 
So, from the standpoint of our texture (2), the estimated not full maximality 
of solar $\nu_e \rightarrow \nu_e$ oscillations may be considered as an 
argument for the existence of the LSND effect.

The final results (14), (18) and (22) follow from the first three oscillation 
formulae (11), if {\it either}

%Eq. 25
\begin{equation} 
 m^2_4 \ll m^2_1 \simeq m^2_2 \simeq m^2_3 
\end{equation} 

\ni with

%Eq. 26
\begin{equation} 
m^2_1 \sim 1\;\;{\rm eV}^2\;\;,\;\; m^2_4 \ll 1\;\;{\rm eV}^2\;\;,\;\; 
\Delta m^2_{21} \sim 10^{-7}\;{\rm eV}^2 \;\;,\;\;  \Delta m^2_{32} 
\sim 3\times 10^{-3} \;{\rm eV}^2 
\end{equation} 

\ni {\it or}

%Eq. 27
\begin{equation} 
m^2_1 \simeq m^2_2 \ll m^2_3 \ll m^2_4 
\end{equation} 

\ni with

%Eq. 28
\begin{equation} 
m^2_1 \ll 1\;\;{\rm eV}^2\;\;,\;\; m^2_4 \sim 1\;\;{\rm eV}^2\;\;,\;\; 
\Delta m^2_{21} \sim 10^{-7} \;{\rm eV}^2 \;\;,\;\; \Delta m^2_{32} 
\sim 3\times 10^{-3} \;{\rm eV}^2 \;.
\end{equation} 

\ni In both cases $\Delta m^2_{21} \ll \Delta m^2_{32} \ll |\Delta m^2_{41}| 
\sim 1\;{\rm eV}^2 $. The first case of $ m^2_4 \ll m^2_1 \sim 1 \;{\rm eV}^2 
$, where the neutrino mass state $ \nu_4$ induced by the sterile neutrino 
$\nu_s$ gets a vanishing mass, seems to be more natural than the second case 
of $ m_3^2 \ll m^2_4 \sim 1 \;{\rm eV}^2 $, where such a state gains a 
considerable amount of mass $\sim 1 \;{\rm eV} $ "for nothing". This is 
so, unless one believes in the liberal maxim "whatever is not forbidden is 
allowed". Note that in the first case the neutrino mass states 
$ \nu_1\,, \,\nu_2\,, \,\nu_3 $ get their considerable masses 
$\sim 1 \;{\rm eV}$ through spontaneously breaking the electroweak 
SU(2)$_L \times$U(1) symmetry which, if it were not broken, would forbid 
these masses.

Finally, for the Chooz reactor experiment [5], where it happens that 
$(x_{ji})_{\rm Chooz} \simeq (x_{ji})_{\rm atm}$, the first Eq. (11) predicts 

%Eq. 29
\begin{equation} 
P(\bar{\nu}_e \rightarrow\, \bar{\nu}_e)_{\rm Chooz}  \simeq  
P( \bar{\nu}_e \rightarrow \bar{\nu}_e)_{\rm atm} 
\simeq 1 -  \frac{1}{2} (1 + c^2_{14}) s^2_{14}   
\end{equation} 

\ni with $\frac{1}{2}(1 + c^2_{14})s^2_{14} \sim 0.095$. In terms of the 
usual two--flavor formula, the negative result of Chooz experiment excludes 
the disappearance process of reactor $ \bar{\nu}_e$'s for moving 
$\sin^2 2\theta_{\rm Chooz} \stackrel{>}{\sim} 0.1 $ and 
$\Delta m^2_{\rm Chooz} \stackrel{>}{\sim} 3\times 10^{-3}\;{\rm eV}^2 $. 
In our case $\sin^2 2\theta_{\rm Chooz} = \frac{1}{2} 
(1 + c^2_{14})s^2_{14}\sim 0.095$ for $\sin^2 x_{\rm Chooz} \sim 1$. Thus, 
the Chooz effect for reactor $\bar{\nu}_e$'s may appear at the edge (if only 
the LSND effect exists with $\sin^2 2\theta_{\rm LSND} = 
\frac{1}{2} s^4_{14} \sim 5\times 10^{-3}$). 

From the neutrinoless double $\beta$ decay, not observed so far,  the 
experimental bound $ \overline{M}_{ee} \equiv |\sum_i U^2_{e i} m_i| 
\stackrel{<}{\sim}$ [0.4 (0.2)---$1.0 (0.6)]$ eV follows [9] (here, 
$U^2_{ei}$ appears even if $U^*_{ei} \neq U_{ei}$). On the other hand, 
with the values  $c^2_{14} \sim 0.9$ and $s^2_{14} \sim 0.1$ the first 
Eq. (5) gives

%Eq. 30
\begin{equation} 
\overline{M}_{ee} = |{M}_{ee} |\sim \frac{1}{2}|0.9 m_1 + 0.1 m_4 + m_2|\;,
\end{equation}

\ni what in the case of Eq. (25) with $m_1 \!\sim\! \pm1$ eV and 
$ m_2 \!\sim\! 1$ eV or Eq. (27) with $ |m_4| \!\sim\! 1$~eV  leads 
to the estimation $\overline{M}_{ee} \sim (0.95, 0.05)$ eV or 
$\overline{M}_{ee} \sim 0,05$ eV, respectively (putting 
$\overline{M}_{ee} = |M_{ee}|$ in Eq. (30) we ignore a tiny --- as 
we believe --- violation of CP: we get $U^*_{ei} = U_{ei}$, since 
$ M_{ee} = \sum_i |U_{e i}|^2 m_i $).

\vfill\eject

\end{document}